\documentclass[sn-mathphys]{sn-jnl}
\usepackage{bm}
\usepackage{bbm}
\usepackage{bbold}
\usepackage{mathtools}
 
\usepackage{graphicx,wrapfig}
\usepackage[utf8]{inputenc}
\usepackage{multicol} 
\usepackage{amssymb}
\usepackage[english]{babel}
\usepackage{color, colortbl}
\definecolor{darkGray}{gray}{0.8}
\usepackage{epsfig}

\usepackage{array, tabularx, multirow, booktabs, diagbox}
\usepackage{makecell}
\setcellgapes{2pt}

\usepackage{placeins}
\usepackage{tabularx}
\usepackage{booktabs}
\usepackage{comment}
 
\jyear{2021}%

\theoremstyle{thmstyleone}%
\theoremstyle{thmstyletwo}%

\theoremstyle{thmstylethree}%

\begin{document}

\title[A stochastic model of cell proliferation and death]{A stochastic model of cell proliferation and death across a sequence of compartments}


\author[1,5]{\fnm{Hanan} \sur{Dreiwi}}\email{H.Dreiwi@leeds.ac.uk}
\equalcont{These authors contributed equally to this work.}

\author[1,2]{\fnm{Flavia} \sur{Feliciangeli}}\email{mmff@leeds.ac.uk}
\equalcont{These authors contributed equally to this work.}

\author[3]{\fnm{Mario} \sur{Castro}}\email{marioc@iit.comillas.edu}

\author[1]{\fnm{Grant} \sur{Lythe}}\email{grant@maths.leeds.ac.uk}

\author*[1,4]{\fnm{Carmen} \sur{Molina-Par\'is}}\email{molina-paris@lanl.gov}

\author*[1]{\fnm{Mart\'in} \sur{L\'opez-Garc\'ia}}\email{M.LopezGarcia@leeds.ac.uk}

\affil[1]{\orgdiv{Department of Applied Mathematics, School of Mathematics}, \orgname{University of Leeds}, \orgaddress{\city{Leeds}, \postcode{LS2 9JT}, \country{UK}}}

\affil[2]{\orgdiv{Research $\&$ Development, Pharmaceuticals Systems Pharmacology $\&$ Medicine}, \orgname{Bayer AG}, \orgaddress{ \city{Leverkusen}, \postcode{51368}, \country{Germany}}}

\affil[3]{\orgdiv{Instituto de Investigaci\'on Tecnol\'ogica (IIT)}, \orgname{Universidad Pontificia Comillas}, \orgaddress{\city{Madrid}, \country{Spain}}}

\affil[4]{\orgdiv{T-6 Theoretical Biology and Biophysics, Theoretical Division}, \orgname{Los Alamos National Laboratory}, \orgaddress{\city{Los Alamos}, \state{NM}, \country{USA}}}

\affil[5]{\orgdiv{Department of Mathematics}, \orgname{University of Benghazi}, \orgaddress{\city{Benghazi},  \country{Libya}}}


\abstract{

Cells of the human body have nearly identical genome but exhibit very different phenotypes that allow them to carry out specific functions and react to changes in their surrounding environment.
This division of labour is achieved by cellular division and cellular
differentiation, events which lead to a population of cells
with unique characteristics.
In this paper, we model the dynamics of cells over time  across a sequence of compartments. Cells within a compartment may represent being at the same spatial location or sharing the same phenotype. In this sequence of compartments, cells can either die, divide or enter an adjacent compartment. We analyse a set of ordinary differential equations to describe the evolution of the average number of cells in each compartment over time. We also focus on the progeny of a founder cell in terms of a stochastic process and analyse several summary statistics to bring insights into the lifespan of a single cell, the number of divisions during its lifespan, and the probability to die in each compartment. Numerical results inspired by cellular immune
processes allow us to illustrate the applicability of our techniques.

}




\maketitle


\section{Introduction}
\label{sec1}

 Cellular proliferation, death and differentiation are crucial and ongoing processes during all stages of life. The most famous example of cell differentiation is embryonic development, where unspecialised 
cells give rise to properly functional and differentiated cells~\cite{Martin1981, Evans1981}. However, this progressive sequence of cell type changes takes place not only in embryo cells but also in adult organisms. Cells gradually acquire a different structure, and this process leads to the emergence of a distinct cell population with specific functions. Some examples are skin cells~\cite{zhang2015modelling}, T~cell development in the thymus~\cite{Sawicka2014}, or the T~cell exhaustion dysfunctional process~\cite{Wherry2015, Chen2019}. The general idea is that starting from a ``precursor'' or ``stem-cell'' population, cells can differentiate,
and thus, change their ``state'' (phenotype) to produce a stable population of differentiated and functional specific cells.

Recent techniques in experimental cell biology have made it possible to track individual cell states and the progeny of a specific cell~\cite{Hodzic2016, Perie2014}. However, we are still far from fully understanding the cellular heterogeneity arising from the development and cell differentiation processes. Thus, mathematical and computational approaches can help answer some open questions regarding the underlying mechanisms of cellular fate. During the last decades, the development of mathematical models in Biology has significantly increased the ability to gain a better quantitative understanding of the dynamics of cell populations over time. Deterministic models, which do not incorporate randomness and are typically easier to analyse, allow one to describe the dynamics over time of many entities, such as animals or cells. However, when looking at a smaller scale, such as when tracking single-cell behaviour, stochastic fluctuations arising from complex cellular interactions and the molecular events which regulate cellular fate must be considered. 

Within the area of stochastic analysis, the theory of branching processes has been widely used in cellular dynamics~\cite{Kimmel2002}. For example, the classic Galton-Watson model, which was originally developed to study the extinction of family surnames~\cite{watson1875probability} has been successfully applied to cellular dynamics~\cite{Kimmel2002}. The theory of branching processes can answer questions related to the limiting behaviour ({\em e.g.,} probability of extinction versus unlimited cellular growth) of cellular populations. A natural generalization is the multi-type branching process, where individuals are not all of the same type. These models can be effectively used to represent cells changing their spatial location~\cite{de2019fate}, or their phenotype~\cite{Nordon20117} over time. These approaches model cellular dynamics across compartments, where cells in the same compartment are assumed to be homogeneous and behave identically. We can highlight here the seminal work by Matis~\cite{Matis1971}, proposing a stochastic compartmental model for cellular dynamics. Deterministic compartmental models are widely used in pharmacodynamics, as well as in Mathematical Ecology~\cite{Matis1994}, Immunology~\cite{Eftimie2016} and Epidemiology~\cite{Capasso2020}.

Here we consider a sequence of compartments and model the process by which, from a stem-cell pool (in compartment $C_1$), cells can undergo differentiation and migration events across adjacent compartments, potentially leading to the rise of a differentiated terminal progeny population (in compartment $C_N$). Cells of each compartment can divide, die or transit to adjacent compartments ({\em e.g.,} representing potentially reversible phenotype change). In our general stochastic model, differentiation can be either considered to be irreversible ({\em e.g.,} cancer cell mutations) or arising from a mutation event that has a negligible probability of being reversed ({\em e.g.,} embryonic cell program). On the other hand, cells are short-lived compared to the host, so a continuous regeneration process is required to maintain cell populations. Recent advances in genetic labelling~\cite{Klein2011} have confirmed the existence of different types of division, which play an essential role in the ability and flexibility of a cell pool to expand (or contract) when required. Different division events can drive different biological processes according to whether the less differentiated pool where the division happens expands to one new cell (self-renewal), stays of the same size (asymmetric division, where one of the daughter cells changes phenotype), or shrinks (symmetric division, where both daughter cells change phenotype)~\cite{Barile2020}. The inclusion of symmetric and asymmetric cell division allows one to analyse how a population of cells needs to be flexible in expanding and how epigenetic information is established and inherited in the cells of a complex and multi-cellular organism.

 In Section~\ref{sec_gen_model} we describe the dynamics of cells dividing, dying or exiting across a sequence of compartments over time, in terms of a continuous-time Markov chain. The mean behaviour of the system is analysed in Section~\ref{sec_num_cell}. In Section~\ref{sec_genealogy} we focus on studying the proliferative potential of the system at the population level by quantifying the number of cells within the genealogy of a single progenitor cell. In both  sections, we consider either the situation where differentiation events can be reversible, or a mathematically more simple ``irreversible model'', where differentiation to the next compartment cannot be reversed. In Section~\ref{sec_singleCell} our focus is on investigating summary statistics related to a single cell which we track over time. We conclude, in Section~\ref{sec_results}, by proposing a set of numerical results inspired by cellular immune processes in order to illustrate our approach.

\section{Stochastic model}
\label{sec_gen_model}

We propose a stochastic model of cell proliferation, death and differentiation (or migration) across an ordered sequence of compartments. Cells in a given compartment may represent co-location in the same spatial compartment, or belong to a common phenotype, or share some common characteristics. We consider a sequence of compartments $C_i$, $i\in\{1,\dots,N\}$, which cells follow, behaving independently of each other as typically considered within the theory of continuous-time branching processes~\cite{Kimmel2002}. 

We start with a general stochastic model in this section, and consider  
cellular events, inspired by 
 some recent mathematical models~\cite{Barile2020,Sawicka2014,zhang2015modelling,de2019fate}.
 We assume that each of these events
 takes place at a certain per cell rate. Particular situations of interest arise from setting some of these rates equal to zero, so that some events are not allowed to happen, as we illustrate in our case studies in Section~\ref{sec_results}.  Each cell in a given compartment, $C_i$, can divide, die or exit to one of the two adjacent compartments. When a division occurs, daughter cells might both belong to the same compartment as the mother (typically referred to as {\it self-renewal}), both daughter cells might both instantaneously belong to the next compartment ({\it symmetric division}), or one daughter cell might belong to the same compartment as the mother while the other belongs to the next compartment ({\it asymmetric division})~\cite{Barile2020,zhang2015modelling}. 

\begin{figure}[htp!]
\centering
\includegraphics[scale=1.85]{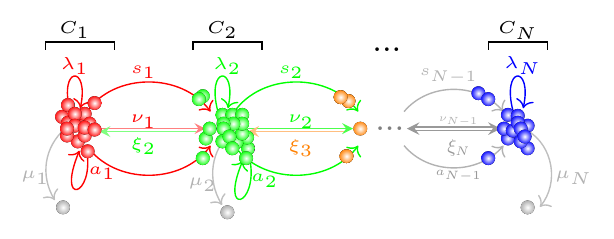}
\caption{Schematic representation of the general stochastic model for cell proliferation, death and differentiation across an ordered sequence of compartments.}
\label{fig:GeneralModel}
\end{figure}

 Our stochastic model, depicted in Figure \ref{fig:GeneralModel}, can be described in terms of the continuous-time Markov chain (CTMC) ${\cal X}=\{({\bm C}_1(t),{\bm C}_2(t),\dots,{\bm C}_N(t)):\ t\geq0\}$, where ${\bm C}_i(t)$ represents the number of cells in compartment $C_i$ at time $t$, with state space given by ${\cal S}=\{0,1,2,\dots\}^N=\mathbb{N}_0^N$. Cellular events, labelled E1 to E5, represent transitions across states $(n_1,\dots,n_N)\in{\cal S}$ as follows:

\begin{itemize}

    \item[(E1)] Self-renewal (cellular division where both daughter cells remain in the same compartment as the mother) can occur in any compartment $C_i$, with per cell rate $\lambda_i$, for $i\in\{1,\dots,N\}$,
    \begin{eqnarray*}
    (n_1,\dots,n_{i-1},n_i,n_{i+1},\dots,n_N) &\xrightarrow{\lambda_in_i}& (n_1,\dots,n_{i-1},n_i+1,n_{i+1},\dots,n_N).
    \end{eqnarray*}

    \item[(E2)]  Asymmetric division (cellular division where one of the daughter cells remains in the same compartment as the mother, while the other goes to the next compartment) occurs in compartment $C_i$ with per cell rate $a_i$,  for $i\in\{1,\dots,N\}$,
    \begin{eqnarray*}
    (n_1,\dots,n_{i-1},n_i,n_{i+1},\dots,n_N) &\xrightarrow{a_in_i}& (n_1,\dots,n_{i-1},n_i,n_{i+1}+1,\dots,n_N).
    \end{eqnarray*}
    
    \item[(E3)] Symmetric division (cellular division where both daughter cells instantaneously move to the next compartment) occurs in compartment $C_i$ with per cell rate $s_i$,  for $i\in\{1,\dots,N\}$,
    \begin{eqnarray*}
    (n_1,\dots,n_{i-1},n_i,n_{i+1},\dots,n_N) &\xrightarrow{s_in_i}& (n_1,\dots,n_{i-1},n_i-1,n_{i+1}+2,\dots,n_N).
    \end{eqnarray*}
    
    \item[(E4)] Differentiation (or migration) between adjacent compartments can occur with  rates $(\nu_i,\xi_i)$,   for $i\in\{1,\dots,N\}$,
    \begin{eqnarray*}
    (n_1,\dots,n_{i-1},n_i,n_{i+1},\dots,n_N) &\xrightarrow{\nu_in_i}& (n_1,\dots,n_{i-1},n_i-1,n_{i+1}+1,\dots,n_N),\\
    (n_1,\dots,n_{i-1},n_i,n_{i+1},\dots,n_N) &\xrightarrow{\xi_in_i}& (n_1,\dots,n_{i-1}+1,n_i-1,n_{i+1},\dots,n_N),
    \end{eqnarray*}
   with $\nu_N=\xi_1=0$.
   
    \item[(E5)] Cells can die in any compartment $C_i$ with per cell rate $\mu_i$, $i\in\{1,\dots,N\}$,
    \begin{eqnarray*}
    (n_1,\dots,n_{i-1},n_i,n_{i+1},\dots,n_N) &\xrightarrow{\mu_in_i}& (n_1,\dots,n_{i-1},n_i-1,n_{i+1},\dots,n_N).
    \end{eqnarray*}

\end{itemize}

 We note that cells in the last compartment $C_N$ cannot symmetrically or asymmetrically divide, or differentiate.

\subsection{Number of cells in each compartment}
\label{sec_num_cell}

We first describe the dynamics of the process by focusing on the time course of the average numbers of cells, 
$ \mathbb{E}[{\bm C}_{i}(t)]$, that obey the following system of differential equations~\cite{Matis1971} 
    \begin{eqnarray}
        \frac{d\, \mathbb{E}[{\bm C}_{1}(t)]}{dt} &=& -(\mu_1+\nu_1+s_1-\lambda_1) \mathbb{E}[{\bm C}_{1}(t)] + \xi_2\mathbb{E}[{\bm C}_{2}(t)],\nonumber\\
        \frac{d\, \mathbb{E}[{\bm C}_{i}(t)]}{dt}& =&  (\nu_{i-1}+a_{i-1}+2s_{i-1})\,\mathbb{E}[{\bm C}_{i-1}(t)] -  (\mu_i+\nu_i+\xi_i+s_i-\lambda_i) \, \mathbb{E}[{\bm C}_{i}(t)]\nonumber \\
        &+& \xi_{i+1}\, \mathbb{E}[{\bm C}_{i+1}(t)], \,\,\,\,\ \quad\quad\quad\quad\quad\quad i \in\{2,...,N-1\},   \\
        \frac{d\,  \mathbb{E}[{\bm C}_{N}(t)]}{dt}& =& (\nu_{N-1}+a_{N-1}+2s_{N-1})\, \mathbb{E}[{\bm C}_{N-1}(t)]- (\mu_N+\xi_N-\lambda_N) \mathbb{E}[{\bm C}_{N}(t)],\nonumber
        \label{general_comp_ODEsys}
    \end{eqnarray}
 where $\mathbb{E}[{\bm C}_i(t)]$ represents the expectation of the random variable ${\bm C}_i(t)$. These equations constitute a homogeneous first-order linear system of ODEs with constant coefficients, which can be written more succinctly in matrix form as  follows
\begin{equation}
\frac{d{\boldsymbol \Phi}}{dt} = {\bf A} {\boldsymbol \Phi},
\label{ph}
\end{equation} 
where 
\begin{equation}
{\boldsymbol \Phi} = \left(\begin{array}{c}  \mathbb{E}[{\bm C}_{1}(t)]\\
\mathbb{E}[{\bm C}_{2}(t)]\\
\vdots\\   
\mathbb{E}[{\bm C}_{N-1}(t)]\\
\mathbb{E}[{\bm C}_{N}(t)]\end{array}\right),\quad 
{\bf A} = \left(
\begin{array}{ccccc}
 -\Delta_1 & \xi_{2} & 0       & \cdots  & 0\\
  \Lambda_1 & -\Delta_2     & \xi_{3}&    \dots     &\vdots\\
  0   &\ddots    & \ddots  &\ddots   &0\\
\vdots&     \vdots     & \Lambda_{N-2} &  -\Delta_{N-1}& \xi_{N}\\
  0   &  \cdots  & 0       & \Lambda_{N-1}    &-\Delta_N
\end{array}
\right),
\label{A1}
\end{equation}
 with 
\begin{eqnarray*}
\Delta_1 &=& \mu_1+\nu_1+ s_1 -\lambda_1,\\
\Delta_i &=& \mu_i+\nu_i+s_i+ \xi_{i}-\lambda_i, \quad \Lambda_{i-1} \ =\  \nu_{i-1}+a_{i-1}+2s_{i-1},\ i\in\{2,\dots,N-1\},\\
\Delta_N &=& \mu_N+ \xi_{N} - \lambda_N,\quad \Lambda_{N-1} \ =\  \nu_{N-1}+a_{N-1}+2s_{N-1}.
\end{eqnarray*}
The initial value problem in \eqref{ph} with ${\bf \Phi}_0 = {\bf \Phi}(0)$ has a unique solution~\cite[Theorem 4.1]{allen2007introduction} 
\begin{equation}
 {\boldsymbol \Phi}(t) = e^{{\bf A}t}{\boldsymbol \Phi}_0,  
\label{ph_sol}
\end{equation} 
where $e^{{\bf A}t}$ represents the matrix exponential,  
$$
e^{{\bf A}t} ={\bf I}+{\bf A}t+{\bf A}^2 \frac{t^2}{2!}+{\bf A}^3 \frac{t^3}{3!}+...= \sum_{i=0}^{+\infty}\frac{({\bf A}t)^i}{i!}.
$$
The system of equations \eqref{ph} admits $\lim_{t\rightarrow + \infty}{\boldsymbol \Phi}(t)={\bf 0}_N$ (column vector of zeros) as an asymptotic solution; this is exponential stable (all solutions of the system from any initial conditions converge to 0) if and only if each eigenvalue of ${\bf A}$ has a negative real part~\cite{Callier1991linear}. 

\begin{figure}[htp!]
\begin{center}
\includegraphics[scale=2]{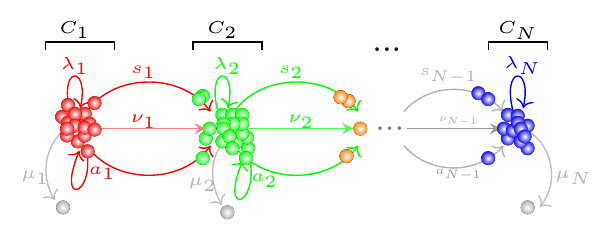}
\caption{Schematic diagram of the general compartmental model without reverse differentiation events.}
\label{Fig:IrreversibleModel}
\end{center}
\end{figure}  

 We note that, in many biological applications, some of the parameters in Fig.~\ref{fig:GeneralModel} will be zero,
 and thus, the analysis of such systems
 will be much simpler. 
 For instance,  when differentiation is irreversible~\cite{Sawicka2014,Barile2020}, so that $\xi_i=0$ for $i\in\{2,\dots,N\}$ as shown in
 Figure~\ref{Fig:IrreversibleModel}. We will refer to this situation as the {\it irreversible model}. In this case, and if one focuses on a single cell starting in compartment $C_1$ at time $t=0$, ${\boldsymbol \Phi}(0)=(1,0,\dots,0)^T$, it is possible to get the average behaviour of the system in terms of the following explicit expressions
\begin{eqnarray}
\mathbb{E}[{\bm C}_{i}(t)] &=& \left\{\begin{array}{ll}
 e^{-\Delta_it}, & i=1,\\
\left( \prod\limits_{l=1}^{i-1} \Lambda_l \right)  \sum\limits_{j=1}^i e^{-\Delta_j t}   \prod\limits_{\substack{m=1 \\ m\neq j}}^i (\Delta_m-\Delta_j)^{-1}, &  i\in\{2,\dots,N\}.
\end{array}\right.
\label{recursive_sol_ncomp_no_fd}
\end{eqnarray}
The expression above is only well-defined if $\Delta_i\neq \Delta_j$ for all pairs $(i,j)$. 
If this is not the case, alternative similar explicit solutions can be found. For example, 
when $\Delta_i=\Delta_j$ for all $i,j\in \{2,\dots,N\}$, then the formula above simplifies to
\begin{equation}
    \mathbb{E}[{\bm C}_{i}(t)] = \left( \prod\limits_{l=1}^{i-1} \Lambda_l \right) \frac{t^i}{i!}e^{-\Delta_it}.
    \label{means_simple}
\end{equation}
 From  the expressions above, it is clear that in the irreversible model $\lim_{t\rightarrow+\infty}\mathbb{E}[{\bm C}_i(t)]=0$ when $\Delta_i>0$, $\forall i\in\{1,\dots,N\}$. This is consistent since the parameters $-\Delta_i$ are the eigenvalues of ${\bf A}$ in this case. 
 For the applications we have in mind~\cite{Sawicka2014,zhang2015modelling}, it is of interest to quantify the cumulative number of cells, on average, that arrive to the final compartment $C_N$ starting with a single cell in compartment $C_1$. To answer this question, one can set $\lambda_N=\mu_N=0$, so that cells which arrive to $C_N$ accumulate and can be counted, and then compute in the irreversible model, from the expression above
  $$ \lim_{t\rightarrow+\infty} \mathbb{E}[ {\bm C}_{N}(t)]=\prod_{i=1}^{N-1}\frac{\Lambda_i}{\Delta_i}.$$

 A particular feature of this system is that cells behave independently from each other. This means that the dynamics of the genealogy of a set of $M$ precursor 
 (or progenitor cells)  in compartment $C_1$ at time $t=0$, can be analysed as $M$ independent stochastic processes. Thus, in Section~\ref{sec_genealogy} we focus on analysing a number of summary statistics of interest which are related to the genealogy of a single cell starting in a given compartment (typically compartment $C_1$).

\subsection{The genealogy of a single precursor cell}
\label{sec_genealogy}

\begin{figure}
\begin{center}
\includegraphics[scale=0.7]{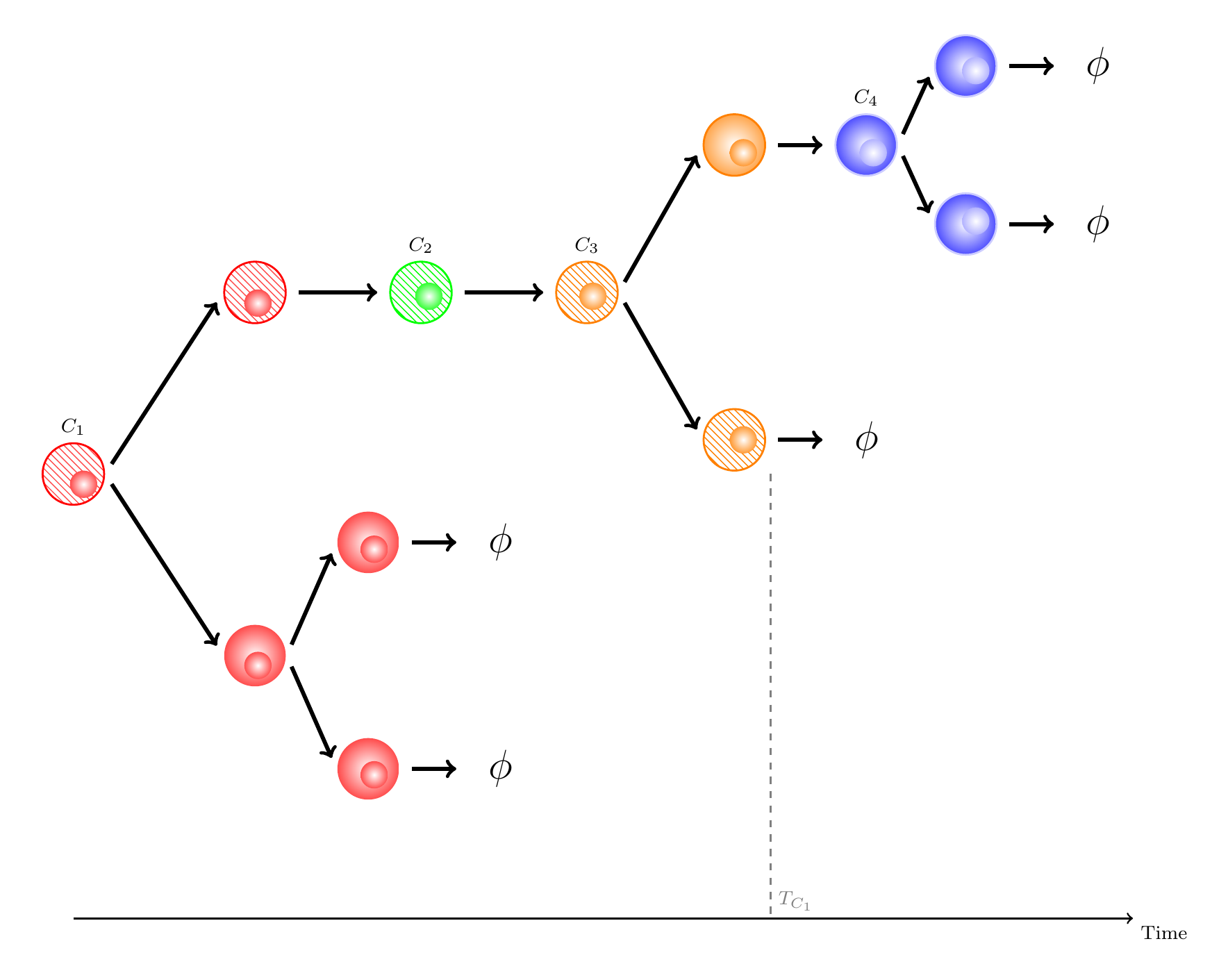} 
 \caption{A realisation of the stochastic process following the genealogy of a single precursor cell which starts in compartment $C_1$. The cell tracked (see Section~\ref{sec_singleCell}) is depicted as striped, and where the colour indicates the compartment where it is at any given time. Here, the tracked cell dies  in $C_3$ (brown), while its genealogy continues up to $C_4$. In this example, $G_1 = 8 = G_1(1)+G_1(2)+G_1(3)+G_1(4)=4+0+2+2$.}
\label{SCell}
\end{center}
\end{figure}

 We consider a cell starting in a given compartment $C_i$, and we define $G_i$, the random variable representing the total number of cells in the genealogy of this cell. Cells in the genealogy are the daughters, granddaughters, \ldots, of the precursor cell, which arise from division events (either self-renewal, asymmetric or symmetric) in any compartment over time, not including the precursor cell. It is a summary statistic of the process which quantifies the proliferative potential of a single cell. For example, in Figure~\ref{SCell}, we represent a particular realisation of the stochastic process, where $G_1=8$. 
 
 The mean number of cells in the genealogy of a precursor cell, $m_i=\mathbb{E}[G_i]$, for any initial compartment of interest $i\in\{1,\dots,N\}$, can be obtained via first-step arguments by conditioning on the next event that occurs in the stochastic process. This approach leads to the following system of equations
\begin{eqnarray*}
    \Delta_1m_1 &=& \Lambda_1m_2 + 2(\lambda_1+a_1+s_1),\nonumber \\
    \Delta_im_i &=& \Lambda_i m_{i+1} +\xi_{i} m_{i-1} +2(\lambda_i+a_i+s_i),\quad i\in\{2,\dots,N-1\}, \\
    \Delta_Nm_N &=& \xi_N m_{N-1} + 2\lambda_N. \nonumber
\end{eqnarray*}
The system above can be expressed in matrix form via the column vectors ${\bf m}=\left(m_1, \dots, m_N \right)^T$ and $ {\bf b}=\left( 2(\lambda_1+a_1+s_1),\dots, 2(\lambda_{N-1}+a_{N-1}+s_{N-1}),\right.$ $\left.2\lambda_N\right)^T$ as follows
\begin{equation}
 {\bf J}\, {\bf m} = {\bf b},
\label{sys2}
\end{equation}
with a tri-diagonal coefficient matrix
\begin{equation}
{\bf J}= \left(
\begin{array}{cccccc}
 \Delta_1&-  \Lambda_1 & 0    & 0   & \cdots  & 0\\
  -  \xi_{2}  & \Delta_2     & -\Lambda_2&   0&  \cdots     &0\\
  0 & -\xi_3 & \Delta_3&   -\Lambda_3&   \cdots    &0\\
  \vdots   &\ddots  &\ddots  & \ddots  &\ddots   & \vdots \\
0&   \cdots &  0    &- \xi_{N-1} & \Delta_{N-1} & -\Lambda_{N-1}\\
  0   &  \cdots  & 0  &0     & - \xi_{N}    & \Delta_{N} 
\end{array}
\right).
\label{J1}
\end{equation}

 One can exploit the tri-diagonal structure of ${\bf J}$ to get an explicit or recursive solution of this system. In particular, by following a Gaussian forward-elimination backward-substitution approach, such as the Thomas algorithm~\cite{conte1972elementary,thomas1949elliptic}, one can obtain the recursive equations
\begin{eqnarray}
m_N &=& \rho_N,\quad m_i \ =\ \rho_i - \gamma_i m_{i+1},\quad i\in\{1,\dots,N-1\},
\label{re_1}
\end{eqnarray}
 where $\gamma_1 = -\Delta^{-1}_1 \Lambda_1$, $\rho_1 = 2\Delta^{-1}_1(\lambda_1+s_1+a_1)$, and
\begin{eqnarray*}
\gamma_i &=& -\frac{\Lambda_i}{\Delta_i+\xi_{i} \gamma_{i-1} },\quad i \in\{2,\dots,N-1\}, \\
\rho_i &=& \frac{2(\lambda_i+s_i+a_i) + \xi_{i} \rho_{i-1}}{\Delta_i+\xi_{i} \gamma_{i-1}},\quad  i\in\{2,\dots,N\}.
\end{eqnarray*}
 This recursive scheme leads to the explicit solution
\begin{eqnarray}
 m_i &=& \sum_{j=i}^N\, (-1)^{j-i}\rho_j \, \left(  \prod_{l=i}^{j-1} \gamma_l \right) ,\quad i\in\{1,\dots,N\},
\label{form_g_1}
\end{eqnarray}
 where  $\prod_{l=i}^{i-1} \gamma_l = 1$. A condition on the parameters arises during the implementation of the recursive scheme,
\begin{eqnarray*}
\Delta_i+\xi_i\gamma_{i-1}&>& 0,\quad i\in\{2,\dots,N\},
\end{eqnarray*}
 for the average values $m_1,\dots,m_N$ to be finite and non-negative, for all $i\in\{1,\dots,N\}$. This ensures that the number of cells in the genealogy of the precursor cell is finite with probability one, $\mathbb{P}(G_i<+\infty)=1$, since $m_i=\mathbb{E}[G_i]=\mathbb{E}[G_i \vert G_i<+\infty]\mathbb{P}(G_i<+\infty)+\mathbb{E}[G_i \vert G_i=+\infty]\mathbb{P}(G_i=+\infty)$. 
 

In the irreversible case ---where $\xi_i=0$, $i\in\{2,\dots,N\}$---, the solution above simplifies to 
\begin{eqnarray} 
m_i=2\sum^{N}_{j=i} \frac{\lambda_j+a_j+s_j}{\Delta_{j}}\left(\prod_{l=i}^{j-1}   \frac{\Lambda_l}{\Delta_l}\right),\quad i\in\{1,\dots,N\}, \label{Set2_1}
\end{eqnarray}   
where for $j=N$, we set $a_N=s_N=0$. For $i=N$, the empty product above is equal to one, so that $m_N=\frac{2\lambda_N}{\mu_N-\lambda_N}$. In this  model the condition on the parameters for finite and non-negative solutions becomes $\Delta_i>0$ for all $i\in\{1,\dots,N\}$. This is consistent with direct inspection of  Figure~\ref{Fig:IrreversibleModel}, where $\Delta_i>0$ avoids unlimited accumulation of cells in compartment $C_i$. 

 We note that cells in the genealogy of the precursor cell belong to different compartments, as depicted by the colours in Figure~\ref{SCell}. For a precursor cell starting in compartment $C_i$, those compartments $C_j$, $j\in\{1,2,\dots,N\}$ with more proliferative potential will contribute more to the value of $G_i$. This will mainly depend on the parameters $(\lambda_j,a_j,s_j,\mu_j,\nu_j,\xi_j)$, but also on the number of cells in the genealogy arriving into that compartment (and  which can thus divide). It is of interest then to split $G_i=\sum_{j=1}^NG_i(j)$ in terms of which compartments the progeny belong to, where $G_i(j)$ represents the number of cells that belong to compartment $C_j$ in the genealogy of the precursor cell starting in compartment $i$. For example, for the stochastic realization in Figure~\ref{SCell}, $G_1=G_1(1)+G_1(2)+G_1(3)+G_1(4)=4+2+0+2=8$. 

 One can follow similar arguments to the ones above to compute the mean quantities $m_i(j)=\mathbb{E}[G_i(j)]$. In particular, for an initial compartment $C_i$, a first-step argument yields the equations
\begin{eqnarray*}
\Delta_i m_{i}(i) &=& 2 \lambda_i + a_i +\Lambda_i\, m_{i+1}(i) +\xi_i m_{i-1}(i),\\
\Delta_im_{i}(i+1) &=& 2 s_i + a_i +\Lambda_i\, m_{i+1}(i+1) +\xi_{i}m_{i-1}(i+1),\\
\Delta_im_{i}(j) &=& \Lambda_im_{i+1}(j) +\xi_{i} m_{i-1}(j),\quad  j\notin\{i,i+1\}, 
\end{eqnarray*}
 where we implicitly set $\xi_{1} = 0$ and $m_{N+1}(j) = 0$ for notational convenience in the equations above. Following a similar recursive approach to the one above, one can get for any $j\in\{1,\dots,N\}$,
\begin{eqnarray*}
m_N(j) &=& \rho_N(j),\quad 
m_i(j) \ =\ \rho_i(j) - \gamma_i m_{i+1}(j),\quad i\in\{1,\dots,N-1\},
\end{eqnarray*}
where $\gamma_1 = -\Delta^{-1}_1 \Lambda_1$, $\rho_1(j) = \Delta^{-1}_1 d_{(1,j)}$, 
and 
\begin{eqnarray*}
\gamma_i &=& -\frac{\Lambda_i}{\Delta_i+\xi_{i} \gamma_{i-1} },\quad\quad i \in\{2,\dots,N-1\}, \\
\\
\rho_i(j) &=& \frac{d_{(i,j)}+ \xi_{i} \rho_{i-1}(j)}{\Delta_i+\xi_{i} \gamma_{i-1}},\quad  i,j\in\{2,\dots,N\},
\end{eqnarray*}
 with
\begin{eqnarray*}
d_{(i,j)} &=&
\begin{cases}
    2\lambda_i + a_i  & \quad \text{if} \quad j=i, \\ 
    2s_i+a_i   & \quad \text{if} \quad  j=i+1,\\
    0 &   \quad \text{otherwise.}
\end{cases}
\end{eqnarray*}
This recursive scheme leads to the explicit solution
\begin{equation}
 m_{i}(j) =\sum_{k=i}^N(-1)^{k-i}\rho_ k(j)\, \left(\prod_{p=i}^{k-1} \gamma_p \right).
\label{form_g_2}
\end{equation}
 This expression simplifies further in the irreversible model, where $\xi_i=0$ for all $i\in\{2,\dots,N\}$. In this instance $m_i(j)=0$ whenever $i>j$, and the equations above simplify to
\begin{eqnarray*}
m_{i}(i) &=& \Delta_i^{-1} (2 \lambda_i + a_i ),\\
m_{i}(j) &=& \Delta^{-1}_{j-1} \left(\prod_{p=i}^{j-2}\Delta^{-1}_p  \Lambda_p\right)\left(d(j-1,j)+ d(j,j)\Delta^{-1}_{j} \Lambda_{j-1}\right),\quad j\in\{i+1, ...,N\},
\end{eqnarray*}
 for any $i\in\{1,...,N-2\}$. Here, $m_{N}(N) = \frac{2\lambda_N}{\Delta_N}$.

\subsection{Single-cell analysis}
\label{sec_singleCell}

 We consider here the case where no symmetric or asymmetric division occurs in the irreversible model, $s_i=a_i=\xi_i=0$ for all $i\in\{1,\dots,N\}$. In the previous sections, we have analysed several summary statistics related to the population of cells over time ({\em e.g.,} the genealogy of a precursor cell). In this section, instead, we aim to investigate a number of summary statistics related to the lifespan of a single cell, extending the single-cell analysis originally proposed in Ref.~\cite{de2019fate} which considered T~cells in the periphery. 

\begin{figure}[H]
\begin{center}
\includegraphics[width=0.75\textwidth]{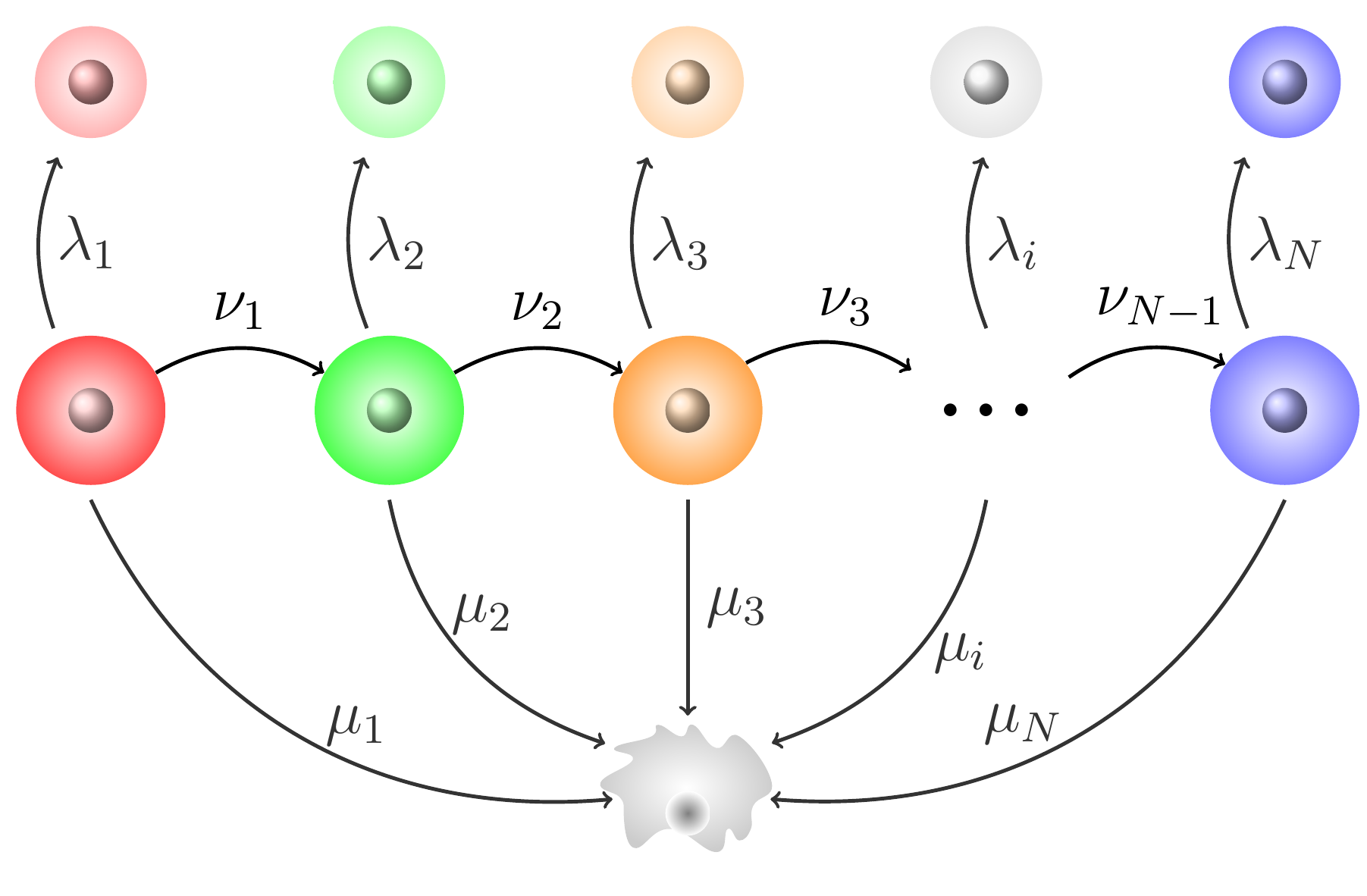} 
\caption{Schematic representation of the process ${\cal Y}$.
}
\label{fig:cell_fate}
\end{center}
\end{figure}

 We propose to track the dynamics of a single cell, starting in compartment $C_i$, until the cell dies. In any given compartment, this cell can divide (and, since both daughters are indistinguishable, we keep tracking one of them), move to the next compartment or die. This dynamics can be represented then in terms of the CTMC ${\cal Y}=\{Y(t):\ t\geq0\}$ over the state space ${\cal S}=\{ C_1, C_2,\dots, C_N,\phi \}$, where $Y(t)$ represents the state of the cell at time $t$: either being in some compartment $C_j$, or having already died (state $\phi$). A schematic representation of the process ${\cal Y}$ is given in Figure~\ref{fig:cell_fate}, while a particular realisation of this stochastic process is depicted in Figure~\ref{Fig:IrreversibleModel}, where the tracked cell is shown as striped.

 We propose to analyse the dynamics of this cellular system by focusing on the following summary statistics:
\begin{itemize}
    \item the lifespan, $T_i$, of the cell, starting in compartment $C_i$,
    \begin{eqnarray*}
    T_i &=& \inf\{t\geq0:\ Y(t)=\phi \vert Y(0)=i\},
    \end{eqnarray*}
 which quantifies the survival potential of cells in the system depending on their initial location;
    \item the number of divisions carried out by the cell during its lifespan, $D_i$, which quantifies the proliferation potential of cells in the system; and
    \item  the probability of this cell dying in each compartment $C_j$; that is, $\beta_i(j)=\mathbb{P}(Y(T_i-\Delta t)=C_j)$ for a small enough $\Delta t$.
\end{itemize}

\subsubsection{Lifespan of a single cell}
\label{sec:lifespan}

 Let $T_i$ be the lifespan of a single cell starting in compartment $C_i$, $i\in\{1,\dots,N\}$, 
\begin{eqnarray*}
T_i &=& \inf\{ t \geq 0: Y(t) = \phi \vert Y(0) = i \}, 
\end{eqnarray*}
 and consider $\tau_i=\mathbb{E}[T_i]$, its average lifespan. By conditioning on the next event that the cell can undergo in this process, one can get the recursive relation
\begin{eqnarray*}
(\mu_i+\nu_i) \tau_{i} &=&  \nu_i \tau_{i+1}  + 1,\quad i\in\{1,\dots,N-1\},
\end{eqnarray*}
\par\noindent with $\tau_{N}=\mu_N^{-1}$. These equations can be explicitly solved as follows
\begin{eqnarray*} 
 \tau_{i} = \sum^{N-i}_{j=0}\frac{1}{\mu_{N-j}+\nu_{N-j}} \left(\prod_{r=i}^{N-(j+1)} \frac{\nu_{r}}{\mu_{r}+\nu_r}\right),\quad i\in\{1,\dots,N-1\}, 
\end{eqnarray*}
 and $\tau_{N} = \mu_{{N}}^{-1}$, since $T_N\sim Exp(\mu_N)$. A similar approach allows one to compute the Laplace-Stieltjes transform of $T_i$, and any of its order moments. For example, the second order moment of the lifespan of a cell starting in compartment $C_i$ is given by 
\begin{eqnarray*}
\mathbb{E} [T_{i}^2] = \sum^{N}_{j=i} R_j \left(\prod_{r=i}^{j-1}\frac{\nu_r}{\mu_r+\nu_r} \right),\quad i\in\{1,\dots,N\},
\end{eqnarray*}
 where 
$R_i= \frac{2(\lambda_i \tau_i+\nu_i \tau_{i+1}+1)}{(\mu_i+\nu_i)(\mu_i + \nu_i +\lambda_i)}$, and $R_N = \frac{2}{\mu_N^2}$.
We note that if the cell starts in the last compartment $C_N$, $\mathbb{E} [T_N^2]=2\mu_N^{-2}$ since $T_N\sim Exp(\mu_N)$. 
 
 \subsubsection{Number of divisions during the lifespan}
 \label{sec:number-divisions-lifespan}
 
 We denote by $D_i$ the number of division events triggered by the tracked cell during its lifespan, starting in compartment $C_i$. One can focus on computing $\eta_i=\mathbb{E}[D_i]$, which by first-step arguments satisfies
 \begin{eqnarray*}
 (\lambda_i+\mu_i+\nu_i)\eta_i &=& \lambda_i(\eta_i+1)+\nu_i\eta_{i+1},\quad i\in\{1,\dots,N-1\},\\
  (\lambda_N+\mu_N)\eta_N &=& \lambda_N(\eta_N+1).
 \end{eqnarray*}
 These equations lead to the explicit solution
\begin{eqnarray*} 
\eta_{i}&=&  \sum^{N-i}_{k=0}\frac{\lambda_{N-k}}{\mu_{N-k}+\nu_{N-k}} \left(\prod_{j=i}^{N-(k+1)} \frac{\nu_{j}}{\mu_j+\nu_j}\right),\quad i\in\{1,\dots,N-1\}.
\end{eqnarray*} 
We note that $\eta_N=\lambda_N\mu_N^{-1}$ since $D_N\sim Geometric (\frac{\lambda_N}{\mu_N+\lambda_N})$. 
These proliferation events can occur at any time during the lifespan of the cell, which is visiting different compartments over time. Thus, one can quantify the proliferation potential of the cell during its eventual visit to each compartment by considering $D_i=\sum_{j=i}^ND_i(j)$, where $D_i(j)$ is the number of divisions which occur exactly in compartment $C_j$ during the lifespan of the cell,
which started in compartment $C_i$. The average values $\eta_i(j)=\mathbb{E}[D_i(j)]$ can be computed similarly, leading to  
\begin{eqnarray*} 
 \eta_i(i) &=&  \frac{\lambda_i}{\mu_i + \nu_i}, \\
\eta_{i}(j) &=& \frac{\lambda_j}{\mu_j+\nu_j}\left(\prod_{k=i}^{j-1} \frac{\nu_{k}}{\mu_k+\nu_k}\right),\quad j \geq i+1.
\end{eqnarray*}
We note that the expression above is
consistent with the interpretation that $D_i(j)\sim Geometric(\frac{\lambda_j}{\lambda_j+\nu_j+\mu_j})$, restricted to the arrival of the cell to compartment $C_j$. In general, we note that
\begin{eqnarray*}
\eta_i(j) &=& \mathbb{E}[D_i(j)]=\mathbb{E}[D_i(j)\ \vert\ \hbox{\it cell ever visits $C_j$}]\mathbb{P}(\hbox{\it cell ever visits $C_j$})\\
&+&\mathbb{E}[D_i(j)\ \vert\ \hbox{\it cell never visits $C_j$}]\mathbb{P}(\hbox{\it cell never visits $C_j$}),
\end{eqnarray*} 
 and, since $\mathbb{E}[D_i(j)\ \vert\ \hbox{\it cell never visits $C_j$}]=0$, the quantity $\eta_i(j)$ also accounts for the probability of the cell not visiting this compartment.

\subsubsection{Cellular fate}
\label{sec:cell-fate}

Finally, the cell will in at a given compartment, and one can compute the probability of this cell dying in each  compartment at the end of its lifespan, which we denote by
\begin{eqnarray*}
\beta_{i}(j) &=& \mathbb{P}(Y(T_i-\Delta t)=C_j),\quad j\geq i,
\end{eqnarray*}
 for small enough $\Delta t$. These probabilities are given by
\begin{eqnarray*}
\beta_{i}(i) &=& \frac{\mu_i}{\mu_{i} + \nu_{i}},\quad\beta_{i}(j) \ =\  \frac{\mu_j}{\mu_j+\nu_j}\prod_{k=i}^{j-1} \frac{\nu_{k}}{\mu_k+\nu_k} \quad j\geq i+1.
\end{eqnarray*}

\section{Results}
\label{sec_results}

We illustrate our approach
with two applications.
In Section~\ref{sec_hypo_scenario}, we implement 
the methods from Section~\ref{sec_num_cell} and Section~\ref{sec_genealogy} to explore the impact of asymmetric and symmetric division events on the cellular dynamics, for the case  $N=4$ compartments.
We carry out some sensitivity analysis on the probabilities of self-renewal, asymmetric and symmetric division events happening.
In Section~\ref{sec_thymus_results} we focus on a particular application of our single-cell analysis from Section~\ref{sec_singleCell} to an existing model of T~cell development in the thymus~\cite{Sawicka2014}.

\subsection{Asymmetric and symmetric division: the
case of four compartments}
\label{sec_hypo_scenario}

We consider here $N=4$ compartments, where the last compartment $C_4$ does not involve any death,
 division or differentiation events, ($\mu_4=\lambda_4=\xi_4=0$) to represent terminal accumulation of cells in it. This allows us to quantify the number of cells that {\it exit} the system formed by the first three compartments (as opposed to dying), which can be of interest in processes such as thymic development~\cite{Sawicka2014}. We select $\mu_i=1.0$ for all $i\in\{1,2,3\}$, so that the unit of time for the rates is the average lifetime of a cell in the system. We aim to analyse the impact of asymmetric and symmetric division on the temporal dynamics, and thus,  set  $\nu_1=\nu_2=\nu_3=0.5$,
in the Irreversible model.

\begin{figure}
    \centering
    \includegraphics[width=\textwidth]{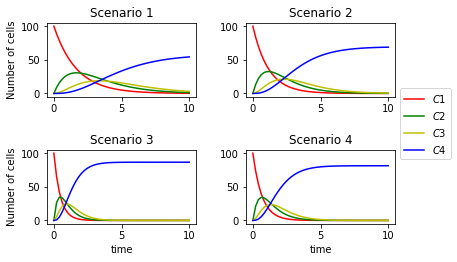}
    \caption{Mean number of cells, $\mathbb{E}[{\bm C}_i(t)]$, predicted over time for compartments $i\in\{1,2,3,4\}$ in scenarios 1, 2, 3 and 4.
    }
    \label{Fig:CaseStudy1_Means}
\end{figure}

 Cells can divide in each compartment at a rate $\omega$, and this division represents self-renewal with probability $p_{SR}$, asymmetric division with probability $p_{AD}$, and symmetric division with probability $p_{SD}$. This is equivalent to setting, with the notation introduced in Section~\ref{sec_gen_model}, $\lambda_i= p_{SR}\cdot \omega$, $s_i=p_{SD}\cdot\omega$ and $a_i=p_{AD}\cdot\omega$, for $i\in\{1,2,3\}$. 
 We choose
  $\omega=0.9$, so that 
the system has significant proliferative potential, and focus on the following scenarios of interest:
\begin{enumerate}[leftmargin=3cm]
    \item[\bf Scenario 1.] $p_{SR}=1.0$, $p_{SD}=0$, $p_{AD}=0$.
    \item[\bf Scenario 2.] $p_{SR}=0.8$, $p_{SD}=0.1$, $p_{AD}=0.1$.
    \item[\bf Scenario 3.] $p_{SR}=0.1$, $p_{SD}=0.8$, $p_{AD}=0.1$.
    \item[\bf Scenario 4.] $p_{SR}=0.1$, $p_{SD}=0.1$, $p_{AD}=0.8$.
\end{enumerate}
Our aim is to explore the impact that asymmetric or symmetric proliferation has on the dynamics of the system (Scenarios 2, 3 and 4), compared to the situation where only self-renewal takes place (Scenario 1). 
 In Figure~\ref{Fig:CaseStudy1_Means} we plot the mean number of cells, $\mathbb{E}[{\bm C}_i(t)]$, in compartments $i\in\{1,2,3,4\}$ in Scenarios 1-4. For compartments $C_i$ with $i\in\{1,2,3\}$, and since $\Delta=\Delta_1=\Delta_2=\Delta_3$ and $\Lambda=\Lambda_1=\Lambda_2=\Lambda_3$, one can directly use Eq. \eqref{means_simple}, while for compartment $C_N$, $N=4$, one has
\begin{eqnarray*}
    \mathbb{E}[{\bm C}_{N}(t)] &=& \left( \frac{\Lambda}{\Delta} \right)^{N-1}-\sum_{j=0}^{N-1} \mathbb{E}({\bm C}_j) \left( \frac{\Lambda}{\Delta} \right)^{N-j}.
\end{eqnarray*}
 In Figure~\ref{Fig:CaseStudy1_Means}, we consider
initial conditions ${\bm C}_1(0)=10^2$, ${\bm C}_2(0)={\bm C}_3(0)={\bm C}_4(0)=0$, representing $10^2$ initial cells in the first compartment and no cells in the other compartments. We observe that an exponential decay in the number of cells in $C_1$ is followed by sequential increases in the other two compartments until a steady-state number of cells is achieved in the {\it exiting} compartment, $C_4$. Interestingly, the dynamics of the system  is faster once symmetric or asymmetric division is considered (Scenarios 2-4 compared to 1),
so that the decay in compartment $C_1$ is quicker and the steady-state is reached sooner.

The fastest dynamics is observed for Scenario 3, where symmetric division is more likely, and the two daughters of a progenitor cell move directly to the next compartment. It is also worth highlighting that symmetric or asymmetric division does not only affect the speed of the dynamics but also the total average number of cells exiting the system ({\it i.e.,} reaching compartment $C_4$), where  $\lim_{t\rightarrow +\infty}\mathbb{E}[{\bm C}_4(t)]$ is significantly larger when asymmetric and (especially) symmetric division can occur. We note here that, importantly, the per cell division rate $\omega$ is equal in all four scenarios. This suggests that, in this type of systems, asymmetric or symmetric division (as opposed to self-renewal proliferation) represents a mechanism which facilitates generating a larger population of exiting cells with the same {\it proliferative effort}. That is, in Scenario 1 where only self-renewal can occur, a larger number of divisions is required in each compartment for enough cells to escape death and differentiate to the next compartment,
and to 
 eventually  reach $C_4$.

\begin{figure}
    \centering
    \includegraphics[width=\textwidth]{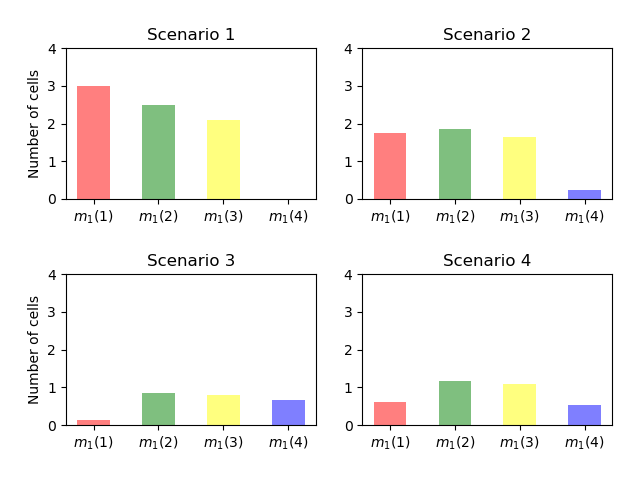}
    \caption{Mean number of cells $m_1(j)$ in the genealogy of a single cell starting in $C_1$, belonging to compartments $C_j$, $j\in\{1,2,3\}$, for Scenarios 1-4.}
    \label{Fig:CaseStudy1_Genealogy}
\end{figure}

 Our comments above are also consistent with 
 the results of Figure~\ref{Fig:CaseStudy1_Genealogy}, where we plot the mean number of cells $m_1(j)$ in the genealogy of a single cell starting in $C_1$, belonging to compartments $C_j$, $j\in\{1,2,3,4\}$, for Scenarios 1-4. In Scenario 1, where only self-renewal can occur, the mean number of cells in compartments $C_1$, $C_2$, $C_3$, and $C_4$ in the genealogy of the progenitor cell decreases monotonically across the sequence of compartments, $m_1(1)>m_1(2)>m_1(3)>m_1(4)$. We note that $m_1(4)=0$ can be explained since it only accounts for progeny cells which arrive into compartment $C_4$ as a direct result of cell proliferation, and no symmetric or asymmetric division is considered in Scenario 1. We also stress here that this monotonic decrease happens even though division and differentiation rates are equal in all compartments $j\in\{1,2,3\}$. This is related to the fact that some cells in the genealogy will die before reaching compartments $C_2$ and/or $C_3$. On the other hand, and related to our comments corresponding to Figure~\ref{Fig:CaseStudy1_Means}, Scenario 1 leads to the largest progeny, suggesting that it is an {\it inefficient} system for reaching a desired population of exiting cells in $C_4$. Asymmetric and (especially) symmetric division events significantly reduce the number of descendants from the single progenitor cell in all compartments, while maximising the number of exiting cells as described by Figure~\ref{Fig:CaseStudy1_Means}. In particular, Scenario 3 is characterised by the highest total average number of cells exiting the  system, $\mathbb{E}[{\bm C}_4(+\infty)]$, as well as the smallest progeny size, $m_1$, while leading to the largest progeny, $m_1(4)$, in the final  compartment. 

\subsection{Tracking a  thymocyte during thymic development}
\label{sec_thymus_results}

\begin{figure}[htp!]
    \centering
    \includegraphics{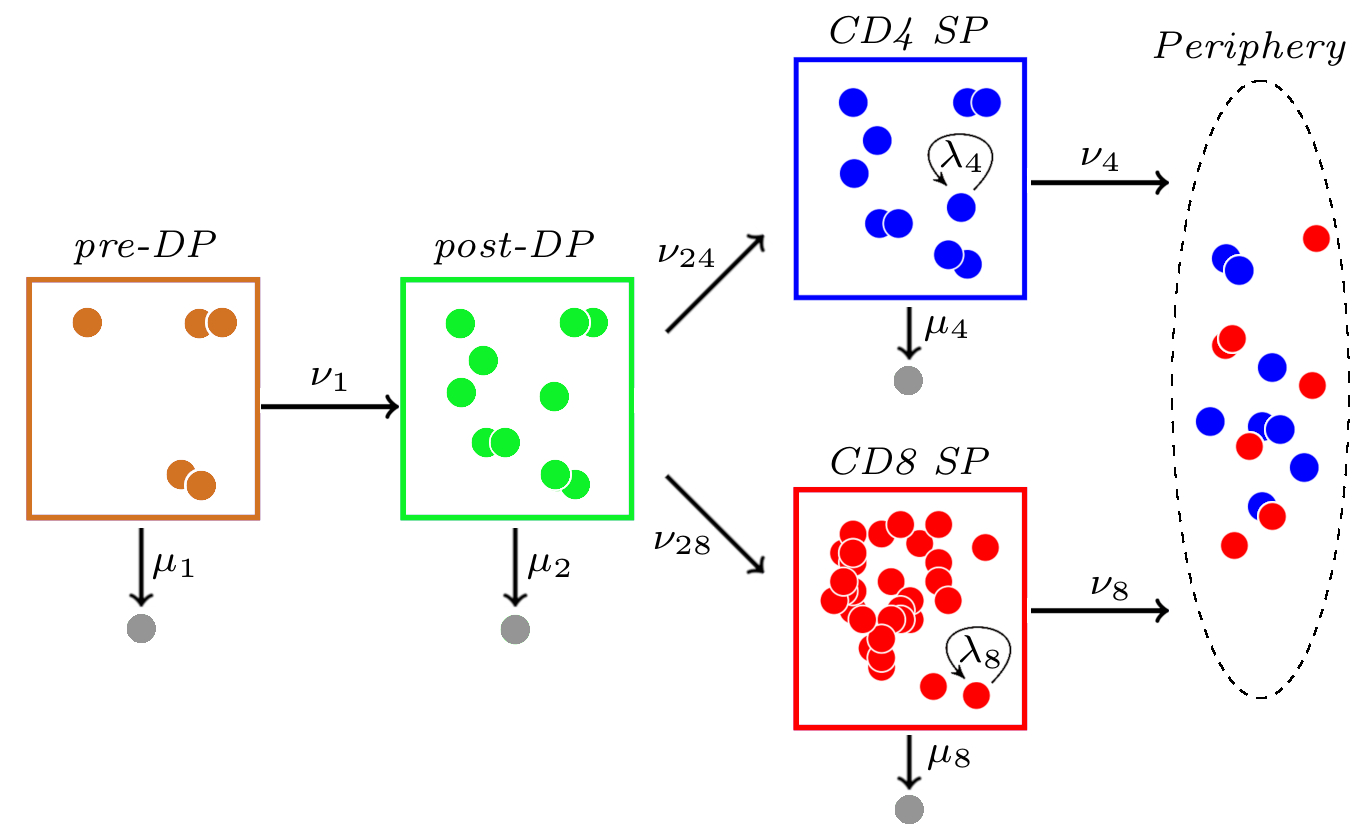}
    \caption{Schematic representation of the thymic development model proposed in Ref.~\cite{Sawicka2014}. Grey cells represent cellular death.}
    \label{Fig:CaseStudy_Thymus}
\end{figure}

We consider now the model proposed in Ref.~\cite[Model 2]{Sawicka2014}, and  shown in Figure~\ref{Fig:CaseStudy_Thymus}. Double negative (DN) thymocytes differentiate to become pre-selection DP thymocytes (pre-DP). In this model, pre-DP is the first compartment, which will contain an initial number of cells (initial condition, $C_1(0)$).
Pre-DPs
 will undergo maturation in the thymus. These cells can progress to the double positive stage (post-DP), where thymocytes express both CD4 and CD8 co-receptors. Post-DP cells that are positively selected transition to the single positive (SP) stage, where they can express either the CD4 or CD8 co-receptor. Some of these cells will then reach the periphery as CD4 or CD8 SP cells. 

We exploit this particular model to illustrate the applicability of our single-cell analysis proposed in Section~\ref{sec_singleCell}. This also allows us to show how our methods can be easily adapted even under slightly modifications to the network topology of compartments (here, a bifurcation in the last compartmental stage is observed, rather than a  linear sequence of compartments). 

\begin{figure}
    \centering
    \includegraphics[width=0.75\textwidth]{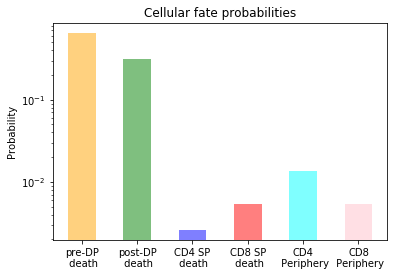}
    \caption{Probabilities of a single  pre-DP cell to die in each of the compartments (pre-DP, post-DP, CD4 SP, or CD8 SP) before reaching the periphery, or reaching the periphery as a CD4 or CD8 SP cell.}
    \label{Fig:CellularFates}
\end{figure}

 First, it is of interest to quantify cellular fate in this system: what is the percentage of pre-DP thymocytes that are predicted to die in each of the compartments during maturation, and what percentage will successfully reach the periphery instead (either as a CD4 or CD8 SP cell). It is clear that our arguments in Section~\ref{sec_singleCell} can be easily adapted to quantify this. In particular, one can slightly redefine the probabilities $\beta_i(j)$ in Section~\ref{sec_singleCell}, focusing on $i=1$ ({\em i.e.,} a single pre-DP thymocyte being tracked), as
\begin{eqnarray*}
\beta_{1}(1) &=& \hbox{\small probability that the pre-DP thymocyte dies in the pre-DP compartment},\\
\beta_{1}(2) &=& \hbox{\small probability that the pre-DP thymocyte dies in the post-DP compartment},\\
\beta_{1}(4) &=& \hbox{\small probability that the pre-DP thymocyte dies in the CD4 SP compartment},\\
\beta_{1}(8) &=& \hbox{\small probability that the pre-DP thymocyte dies in the CD8 SP compartment},\\
\beta_{1}(4P) &=& \hbox{\small probability that the pre-DP thymocyte reaches the periphery as a CD4 SP cell},\\
\beta_{1}(8P) &=& \hbox{\small probability that the pre-DP thymocyte reaches the periphery as a CD8 SP cell}.
\end{eqnarray*}
 Similar solutions to those in Section~\ref{sec_singleCell} can be obtained by incorporating the compartmental bifurcation in the first-step analysis, leading to
\begin{eqnarray*}
\beta_{1}(1) &=& \frac{\mu_1}{\mu_1+\nu_1},\\
\beta_{1}(2) &=& \frac{\nu_1}{\mu_1+\nu_1}\frac{\mu_2}{\mu_2+\nu_{24}+\nu_{28}},\\
\beta_{1}(4) &=& \frac{\nu_1}{\mu_1+\nu_1}\frac{\nu_{24}}{\mu_2+\nu_{24}+\nu_{28}}\frac{\mu_4}{\mu_4+\nu_4},\\
\beta_{1}(8) &=& \frac{\nu_1}{\mu_1+\nu_1}\frac{\nu_{28}}{\mu_2+\nu_{24}+\nu_{28}}\frac{\mu_8}{\mu_8+\nu_8},\\
\beta_{1}(4P) &=& \frac{\nu_1}{\mu_1+\nu_1}\frac{\nu_{24}}{\mu_2+\nu_{24}+\nu_{28}}\frac{\nu_4}{\mu_4+\nu_4},\\
\beta_{1}(8P) &=& \frac{\nu_1}{\mu_1+\nu_1}\frac{\nu_{28}}{\mu_2+\nu_{24}+\nu_{28}}\frac{\nu_8}{\mu_8+\nu_8}.
\end{eqnarray*}
 The proliferation potential of thymocytes during thymic development directly depends on them reaching the CD4 SP and CD8 SP compartments, 
 where they are able to divide, before they reach the periphery. Thus, the average number of divisions initiated by a single pre-DP thymocyte during its thymic development journey, $\eta_1=\eta_1(4)+\eta_1(8)$, can be obtained as 
\begin{eqnarray*}
\eta_1(4) &=& \frac{\nu_1\nu_{24}}{(\mu_1+\nu_1)(\mu_2+\nu_{24}+\nu_{28})}\frac{\lambda_4}{\mu_4+\nu_4},\\
\eta_1(8) &=& \frac{\nu_1\nu_{28}}{(\mu_1+\nu_1)(\mu_2+\nu_{24}+\nu_{28})}\frac{\lambda_8}{\mu_8+\nu_8}.
\end{eqnarray*}
 Finally, the average lifespan of a pre-DP cell during thymic development ({\em }i.e., the mean time until it dies or it reaches the periphery) is given by 
\begin{eqnarray*}
\tau_1 = \frac{1}{\mu_1+\nu_1} \left[ \frac{\nu_1}{\mu_2+\nu_{24}+\nu_{28}}\left( \frac{\nu_{24}}{\mu_4+\nu_4}+\frac{\nu_{28}}{\mu_8+\nu_8}+1\right) +1\right].
\end{eqnarray*}
 We select parameter values from those reported in Ref.~\cite[Section 3.2]{Sawicka2014} (see Table~\ref{Tab:Parameters}). Our methodology allows one to predict that the average lifespan of a pre-DP cell during the maturation process (until it dies or reaches the periphery) is $2.84$ days. During this lifetime, the cell may undergo differentiation and proliferation, before dying in one of the compartments without ever reaching the periphery, or reaching the periphery either as a CD4 or CD8 cell. In Figure \ref{Fig:CellularFates}, we plot the predicted cellular fate probabilities. We observe that the most likely outcomes correspond to cell death, especially during the early stages of maturation (pre-DP and post-DP compartments). Once the cell reaches the CD4 SP compartment, it is more likely for this cell to reach the periphery than dying in that compartment, while these probabilities are comparable in the CD8 case. The average number of proliferation events triggered by this single pre-DP cell is $\eta_1=\eta_1(4)+\eta_1(8)=0.0139+0.0046=0.0185$, meaning that out of $100$ pre-DP cells entering the thymic development process only around $18$ cells are expected to be produced via direct proliferation from the original cells when visiting the CD4 SP or CD8 SP compartments. These small quantities are directly related to the small probabilities of reaching these compartments at all, so that the cell can actually proliferate. Our results here are in agreement with conclusions and insights from \cite{Sawicka2014}, where the authors state that thymic development is an stringent process characterized by an extremely low success rate.

\begin{table}[htp!]
    \centering
    \scalebox{0.9}{
    \begin{tabular}{|c|c|c|c|c|c|c|c|c|c|c|c|}
    \hline
    Rate & $\mu_1$ & $\nu_1$ & $\mu_2$ & $\nu_{24}$ & $\nu_{28}$ & $\lambda_4$
    & $\lambda_8$ & $\mu_4$ & $\mu_8$ & $\nu_4$ & $\nu_8$ 
     \\ \hline
    Value & $0.263$ & $0.137$ & $1.369$ & $0.07$ & $0.054$ & $0.216$ 
     & $0.093$ & $0.04$ & $0.11$ & $0.21$ & $0.14$
 \\ \hline
    \end{tabular}
    }
    \caption{Parameter values as reported in Ref.~\cite[Section 3.2]{Sawicka2014}, in units $days^{-1}$.}
    \label{Tab:Parameters}
\end{table}

\section{Discussion}
\label{sec:discussion}

We have presented a general model for cell proliferation and differentiation in terms of a stochastic process across a sequence of compartments. Cells can divide, die or move to adjacent compartments over time. We have explicitly quantified the average number of cells in each compartment over time under different scenarios of interest ({\em e.g.,} the irreversible model, where differentiation cannot be reversed), and analysed the genealogy of a single progenitor cell in terms of summary statistics appropriately defined. Single-cell analysis allows one to track the journey of one cell during its lifetime across the system of compartments, where we are able to quantify the cell lifetime, its proliferation potential and the likelihood of the different potential cellular fates. We have presented a number of numerical results to illustrate the applicability of our techniques, and the impact that model parameters can have on the corresponding predictions. 

 A particular limitation of our approach is that it relies on cells behaving independently from each other, which is directly related to the fact that the corresponding system of ODEs for the average behaviour of the process is linear. This allows one to implement techniques 
 from the theory of branching processes, and makes the single-cell analysis implementation feasible (where one can follow a single tracked cell while neglecting the dynamics of its neighbours). On the other hand, cellular interactions might not be negligible in some situations ({\em e.g.,} under competition for resources, where logistic growth-type models might be more appropriate). Relaxing this particular assumption is thus, the aim of future work.

\section{Acknowledgments}

H.~Dreiwi was a Daphne Jackson Fellow during the period 2019-2021 and would like to thank the Daphne Jackson Trust and the University of Leeds, School of Mathematics, for their financial and academic support. This work has been supported by the European Commission through the Marie Sklodowska-Curie Action
(H2020-MSCA-ITN-2017) Innovative Training Network Quantitative T~cell Immunology and Immunotherapy (QuanTII), project number 764698 (FF, GL,
and CMP).

\bibliography{sn-bibliography}

\end{document}